
\magnification=\magstep1
\baselineskip=12pt
\hsize 6.0 true in
\vsize 9.0 true in

\voffset = - .2 true in

\font\tentworm=cmr10 scaled \magstep2
\font\tentwobf=cmbx10 scaled \magstep2

\font\tenonerm=cmr10 scaled \magstep1
\font\tenonebf=cmbx10 scaled \magstep1

\font\eightrm=cmr8
\font\eightit=cmti8
\font\eightbf=cmbx8
\font\eightsl=cmsl8
\font\sevensy=cmsy7
\font\sevenm=cmmi7

\font\twelverm=cmr12
\font\twelvebf=cmbx12
\def\subsection #1\par{\noindent {\bf #1} \noindent \rm}

\def\mid {\let\rm=\tenonerm \let\bf=\tenonebf \rm \bf}

\def\para{\par \vskip 12 pt}

\def\head{\let\rm=\tentworm \let\bf=\tentwobf \rm \bf}

\def\heading #1 #2\par{\centerline {\head #1} \smallskip
 \centerline {\head #2} \vskip .15 pt \rm}

\def\eight{\let\rm=\eightrm \let\it=\eightit \let\bf=\eightbf
\let\sl=\eightsl \let\sy=\sevensy \let\m=\sevenm \rm}

\def\foots{\noindent \eight \baselineskip=10 true pt \noindent \rm}
\def\sexion{\let\rm=\twelverm \let\bf=\twelvebf \rm \bf}

\def\section #1 #2\par{\vskip .35true in \noindent {\mid #1} \enspace {\mid #2}
  \para \noindent \rm}

\def\abstract#1\par{\para \foots {\bf Abstract: \enspace}#1 \para}

\def\author#1\par{\centerline {#1} \vskip 0.1 true in \rm}

\def\abstract#1\par{\noindent {\bf Abstract: }#1 \vskip 0.5 true in \rm}

\def\midsection #1\par{\noindent {\sexion #1} \noindent \rm}

\def\sqr#1#2{{\vcenter{\vbox{\hrule height.#2pt
  \hbox {\vrule width.#2pt height#1pt \kern#1pt
  \vrule width.#2pt}
  \hrule height.#2pt}}}}

\pretolerance=10000
\def\n{\noindent}
\def\m{\medskip}
\def\b{\bigskip}
\def\c{\centerline}
\def\n{\noindent}
\def\s{\smallskip}

\line{\hfil IUCAA- 1/93 Jan'93}

\vskip 3cm

\c{\mid Singularity Free Inhomogeneous Cosmological Stiff Fluid Models }
\vskip 1.5 cm

\b

\c{L.K. Patel}
\s
\c{Department of Mathematics, Gujarat University,}
\c{Ahmedabad - 380 009, India}

\vskip 0.5 cm

\c{Naresh Dadhich$^*$}
\c{Inter-University Centre for Astronomy and Astrophysics,}
\c{Post Bag 4, Ganeshkhind, Pune - 411 007, INDIA}

\vskip 2cm

\c{\bf Abstract}

\n We present a singularity free class of inhomogeneous cylindrical universes
 filled
with stiff perfect fluid $(\rho = p)$. Its matter free $ (\rho = 0)$ limit
yield two distinct vacuum spacetimes which can be considered as analogues of
Kasner solution for inhomogeneous singularity free spacetime.
\vskip 1 cm

\n PACS numbers : 04.20.Jb, 98.80. Dr

\vskip 1 cm

\n *E-mail address : naresh@iucaa.ernet.in

\vfill\eject

\n The standard Friedmann-Roberstson-Walker model has been quite successful in
 describing the present state of the Universe. It prescribes homogeneous and
isotropic
distribution for matter. It is though realised that homogeneous and isotropic
character of spacetime
cannot be sustained at all scales, particularly for very early times.
Furthermore,
not to have to assume very special initial conditions as well as for formation
of
large  scale structure in the Universe it is imperative to consider
inhomogeneity
and anisotropy.
\s

\n The first step in this direction came in the form of the study of
anisotropic
 Bianchi models. Then inhomogeneity was also brought in and some inhomogeneous
models were considered [1-5]. One of the main characteristics of the
Einsteinian
cosmology is the
prediction of a big-bang singularity in the finite past. All these models (FRW,
 Bianchi, as well as inhomogeneous) suffer from the singulairty at $t = 0$.
This experience was strongly aided by the general
result that under physically reasonable conditions of positivity of energy,
causality and regularity etc., the initial singularity is inescapable in
cosmology
 so long
as we adhere to Einstein's equations (singularity theorems [6]). This gave rise
to the
folk-lore that the big-bang singularity is the essential property of cosmology
and
it can only be avoided by invoking quantum effects and /or modifying Einstein's
theory.
\s

\n On this background it was really very refreshing when Senovilla [7] obtained
a
new class of exact solutions of Einstein's equations without the big-bang
 singularity. It represented a cylindrically symmetric universe filled with
perfect fluid
$(\rho = 3p)$. It was smooth and regular everywhere, satisfied the energy and
causality conditions, and all the physical as well as geometrical invariants
were
finite for whole of spacetime. This marked the advent of singularity free
cosmology. It
is important to recognise that the occurrence of singularity is not the general
feature of Einstein's equations and for its avoidance it is not always
necessary
to resort to quantum effects. That is the classical Einstein's theory does
admit
cosmological
models without the singularity.

\s

\n One may however wonder how do the sigularity free solutions bypass the
singularity theorems ? It was subsequently shown [8] that they did not obey the
assumption of existence of ``trapped surfaces'' which was never so physically
obvious as other assumptions and had always been a bit of suspect. All causal
curves were shown to be complete demonstrating the absence of singularity.

\s

\n It turns out that inhomogeneous models admit many different kinds of
singularities (not always the spacelike universal big-bang  kind) or none
depending
upon the particular choice of parameters and metric functions [9]. There exists
a general class of singularity free solutions. We have recently generalised the
Senovilla
model [7] to consider viscous fluid [10] and perfect fluid with radial heat
flow
[11]. It was interesting to see that the singularity free character of the
model
remained undisturbed.

\s

\n The matter free $(\rho = p = 0)$ limit of cosmological models is always of
interest. It represents a kind of base spacetime that provides a simple
framework
for analysis of  the basic structure of the model. The de Sitter spacetime  can
be
 taken as a base  (which is truely not empty though $R_{ik} = \wedge g_{ik}$)
for the FRW cosmology while the
Kasner is for homogeneous (Bianchi type I) cosmology. Kasner is perfectly empty
space
($R_{ik} = 0$)
homogeneous cosmological solution. It was very effectively used for the
general
analysis of the initial  big-bang singularity [12].
\s

\n The question we wish to address in this Letter is : Does there exist a
singularity free empty space soltuion and does it arise as the matter free
limit
of a singularity free cosmological model ? In what follows we report a new s
ingularity free inhomogeneous cosmological model describing a perfect fluid
with
the stiff matter equation of
state, $\rho = p$. Its matter free limit $(\rho = p = 0 ) $ gives two distinct
empty space solutions.

\s

\n We take the spacetime to admit two commuting spacelike Killing vectors,
which
are
mutually as well as hypersurface orthogonal (orthogonally transitive $G_2$
cosmologies [13-15]). The solution is given by

\b

\c{$ ds^2 = C^{2(1-b)}{(2mt)} \left[ C^{4b(2b-1)}(mr)  (dt^2 - dr^2 )
-S^2(mr)  C^{2(1-2b)}
(mr)  d\phi^2 \right]$}
$$  - C^{2b}{(2mt)}  C^{4b}{(mr)} dz^2
\eqno(1) $$

\n and

$$ \rho = p = {m^2 \over 2\pi} (b^2 - 1 ) C^{-2(2-b)}(2mt)
 \quad C^{-4b(2b-1)}(mr)
\eqno(2) $$

$$ \theta = 2m (2-b) S(2mt) C^{-2 +b}(2mt)  C^{-2b(2b-1)}(mr) \eqno(3) $$

$$\sigma^2 = {8 \over 3} m^2 (1 - 2b)^2 S^2(2mt) C^{-2(2-b)}(2mt)
C^{-4b(2b-1)}(mt)
\eqno(4) $$

$$ f_r = -2mb (2b-1) S(2mt) C^{-1}(2mt) .\eqno(5) $$

\n Here $m$ and $b$ are free parameters; $\theta, \sigma$ and $f_r$ are
respectively expansion, shear and radial acceleration; and $C(x) = coshx$ and
$S(x) = sinhx$.

\s

\n Clearly the solution is everywhere smooth, regular and finite, and so are
the other
parameters listed above for all values of $t$ and $r$. We have checked that all
the Riemann and Weyl curvatures are finite and regular for whole spacetime.
The solution is new and not covered in the class discussed in Ref.9.  The
energy
condition is obviously satisfied for $b$ taking values outside the range
 $-1 < b <1$.

\s

\n For $\rho > 0$ and decreasing  with $t$ requires $2 >b>1$ or $b < -1$. For
$ b = 2$, $\rho = \rho(r)$ and $\theta = 0$, that means model becomes
stationary
with $\rho$ decreasing with $r$. The $r$-dependence of the parameters has in
general
monotonically decreasing character for all admissible values of $b$. Let us
consider the two cases (i)$ 2>b>1$ and $b< -1$ for     $t-$dependence
separately.

\n Case (i): $2>b>1 .$ $ \rho$ is maximum at $t = 0$ and decreases to zero as
$ t \rightarrow \pm \infty$ while $\theta (t = 0) = 0, \theta (t \rightarrow
 \pm \infty)
\rightarrow  \pm \infty .$ This is quite an acceptable behaviour.
\s

\n  Case (ii): $b < -1. $  $\rho$ has the same behaviour as above while
$\theta (t=0) = 0
, \theta (t \rightarrow  \pm \infty) \rightarrow \pm  0$ , indicating
non-monotonic expansion. It attains the extremum values
for $ S(2mt) = \pm (1 + |b|)^{-1/2}, +$ sign for  $ 0 \leq t \leq \infty$ and
the other for $-\infty \leq t \leq 0$. One would like $\theta$ to be
monotonically
increasing with $t$, which can be provided by the interval between the two
extrema. This however restricts the range of $t$ which is not very nice .
\s

\n Overall the former case may be preferable as it has all through acceptable
behaviour. The parameter $m$ can be chosen as large as we please to make $\rho$
arbitrarily large. Thus the solution may be appropriate for very early times as
it can
provide highly dense state of matter, very rapid expansion and above all no
singularity to worry about.

\s

\n Let us now come to the matter free limit . From (2) it requires
$ b^2 = 1$ leading to two distinct empty space solutions for $b = \pm 1$. Thus
we obtain the singularity free empty space cosmological solutions as follows :

\n For $ b = 1 $

$$ ds^2 = C^4(mr) (dt^2 - dr^2 - C^2(2mt)dz^2 ) - S^2(mr)  C^{-2}(mr) d \phi^2
\eqno(6) $$

\n and for $b = -1 $

$$ ds^2 = C^4(2mt) C^6(mr) \left[ C^6(mr) (dt^2 - dr^2 ) - S^2(mr) d \phi^2
\right] - C^{-4}(mr) C^{-2}(2mt) dz^2 . \eqno(7) $$

\n They have finite and regular Weyl curvatures everywhere and in particular,
$W_{ijke} (b = 1, t=r=0) = -W_{ijke} (b = -1, t = r = 0 ).$ They mark the
boundries of the forbidden range of the parameter $ b~~ (-1 < b < 1)$ for the
above model (1). The solution
(6) is the matter free limit of the case (i) while (7) is for the case (ii)
above.
They are analogues of Kasner solution for singularity free inhomogeneous
spacetime .

\s

\n If we write the metric coefficients in the solutions (6) and (7) as
$C^{2p_i}(2mt)
C^{2q_i}(mr)$ for $ i= 1, ..,4,$ then  $p_i$ and $q_i$ satisfy the Kasnerian
type
relation,

$$ p_1 + p_2 + p_3 - p_4 = 1 = (p_1 + p_2 + p_3 - p_4 )^2 $$

\n and

$$ q_1 + q_2 + q_3 - q_4 = 1 = (q_1 + q_2 + q_3 - q_4 )^2. $$

\n This is quite interesting and perhaps lends strength to the analogy. In the
Kasner case $p_4 = 0$ and there are no $q_i$ as the solution is homogeneous.
We wonder
whether the above relation is a general property of inhomogeneous empty space
cosmological solutions.

\s

\n We have thus presented a physicallyy acceptable singularity free
inhomogeneous
perfect fluid cosmological model with stiff matter equation of state and two
distinct
vacuum solutions as its matter free limit.  Their application to the very early
universe cosmology will be taken up in a future paper. The equation of state
$\rho = p$ is quite appropriate for consideration of a scalar field which has
been very popular with the early Universe cosmologists.
\s

\n Finally we would like to say a word about the occurrence of ``trapped
surfaces''.
It may however seem quite plausible for gravitational collapse but it is not at
all
obvious in the cosmological context. For we have solutions that satisfy all
other
physically pertinent conditions and present quite acceptable cosmological
models.
They seem to always occur in spherically symmetric spacetimes while in
cylindrical
symmetry their occurrence is indefinite. It appears that inhomogeneity and
cylindrical
symmetry seem to be necessary but not sufficient for non-existence of trapped
surfaces.

\vskip 1 cm

\n{\bf Acknowledgement}
\b
\n We thank Tarun Ghosh and Kanti Jotania for help in algebraic computing using
Macsyma and Mathematica.

\vfill\eject

\n {\bf References}
\b

\item{[1]} J. Wainwright and S.W. Goode, Phys. Rev. {\it D 22}, 1906 (1980).
\s
\item{[2]} A. Feinstein and J.M.M. Senovilla, class. Quantum Grav. {\it 6}, L89
(1989).
\s
\item{[3]} W. Davidson, J. Math. Phys. {\it 32}, 1560 (1990).
\s
\item{[4]} L.K. Patel and N. Dadhich, Ap. J. {\it 401}, 433 (1992) .
\s
\item{[5]} L.K. Patel and N. Dadhich, J. Math. Phys. (to appear).
\s
\item{[6]} S.W. Hawking and G.F.R. Ellis, The Large Scale Structure
of the Universe (Cambridge University Press, 1973).
\s
\item{[7]} J.M.M. Senovilla, Phys. Rev. Lett. {\it 64}, 2219 (1990).
\s
\item{[8]} F.J. Chinea, L. Fernandez-Jambrina and J.M.M. Senovilla, Phys. Rev.
{\it D45}, 481 (1992).
\s
\item{[9]} E. Ruiz and J.M.M. Senovilla, Phys. Rev. {\it D45}, 1995 (1992).
\s
\item{[10]}L.K. Patel and N. Dadhich, Phys. Rev. D (Submitted).
\s
\item{[11]}L.K. Patel and N. Dadhich, (To be submitted).
\s
\item{[12]} L.D. Landau and E.M. Lifshitz, The Classical Theory of Fields
(Pergamon Press, New York, 1985).
\s
\item{[13]} C.G. Hewitt and J. Wainwright, Class. Quantum Grav. {\it 7}, 2295
(1990).
\s
\item{[14]} J. Wainwright, J. Phys. {\it A14}, 1131 (1981).
\s
\item{[15]} M. Carmeli, Ch. Charach and S. Malin, Phys. Rep. {\it 76}, 80
(1981).

\bye